\documentclass[twocolumn,showpacs,amsmath,amssymb,prb,apssamp]{revtex4}

\usepackage{graphicx}
\usepackage{dcolumn}
\usepackage{bm}
\usepackage{rotating}
\usepackage{color}

\begin{document}

\preprint{APS/123-QED}

\title{Depinning transition for a screw dislocation in a model solid solution}

\author{S. Patinet\email{sylvain.patinet@cea.fr}
 and L. Proville}
\affiliation{CEA, DEN, Service de Recherches de M\'etallurgie Physique,
F-91191 Gif-sur-Yvette, France }

\date{\today}

\begin{abstract}
On the basis of the classical dislocation theory, the solid
solution hardening (SSH) is commonly ascribed to the pinning
of the edge dislocations.
At the atomic level, the theoretical study of the dislocation cores
contrasts with such a prediction. Using the static molecular simulations
with some interatomic effective potentials, we demonstrate numerically that the
critical resolved shear stress associated with a screw dislocation
in a random Ni(Al) single crystal has same order as the edge
one. Such a result is imposed by the details of the
dislocation stacking fault and the core dissociation into Shockley
partials. The SSH statistical
theory is employed to tentatively predict analytically the data
acquired through our atomistic simulations at different $Al$
concentration.
\end{abstract}

\maketitle

\section{Introduction}
In the alloy manufacturing the solid solution hardening (SSH) is a
standard process which allows to increase the yield stress of a
material by dispersion of some atomic-sized obstacles
across the dislocation glide. The
choice of the impurity and which proportion is required is an
important issue in the commercial alloy
design.\cite{Cahn2001,Sieurin2006} On the condition that these
impurities remain in solution, favored either by a thermal treatment or
the alloy thermodynamics, the microstructure is
unchanged but the dislocations are pinned by the randomly
distributed obstacles.
The dislocation pinning yields an increase
in the material strength without involving large inhomogeneities
such as inclusions or grain boundaries, by contrast to the other methods
as the precipitation strengthening or the strain hardening.

The dislocation
impinging on a random distribution of obstacles is a standard
problem of the theoretical
metallurgy\cite{Mott1948,Mott1952,Nabarro1985,Friedel1964,Haasen1979,Suzuki1991,Butt1980,Chandrasekaran2001}
and the statistical
physics\cite{Labusch1970,Vinokur2002,Brazovskii2004,LeDoussal2000,Rosso2007} as well.
Only recently this problem could have been addressed with
some three-dimensional atomistic simulations
\cite{Rodney2000,Picu2004,Tapasa2004,Rodary2004,Olmsted2005,Bitzek2005,
Marian2006,Curtin2006,Tapasa2006,Proville2006} that shed a new light
on points that were still a matter of debate in material science.
However the difficulty of developing reliable inter-atomic
potentials for modeling dislocations in alloys confines the
atomistic simulations to only few systems. The Ni(Al)
$\gamma$-phase is one of them and corresponds to the
prototypical case for the binary substitutional alloys with a high
order energy and a high solubility limit, about $C_{Al}=10$ at$.$
$\%$ which involves a broad concentration range for the stability of
the solid solution.

The static atomistic
simulations allow to compute the Critical Resolved Shear Stress
(CRSS) for an isolated dislocation in a single crystal of a purely
random solution.\cite{Tapasa2004,Tapasa2006,Proville2006} In the
model Ni(Al) solution, the CRSS associated with an edge dislocation was found\cite{Proville2006}
to increase roughly linearly at a rate of about 30 MPa per atomic percent.
This result agrees with the experimental data on the Ni(Al)
hardness\cite{Cahn2001} ($H$) from which the flow stress $\sigma$
can be deduced by application of the empirical linear
relation\cite{Tabor1951,Gutkin2004} $H= 3 \sigma$ (valid for metallic crystalline materials).
The main question raised by such a
result bears on the role of the screw dislocation segments.
According to a widespread belief drawn on the first order elastic
dislocation theory (see text book as Ref. \onlinecite{Hirth1982}),
the screw dislocation CRSS would be smaller than the edge one.
By contrast, our present atomistic study shows that
the screw CRSS has same order as the edge one for different Al densities
$C_{Al}$, between $2$ at. $\%$ and 12 at. $\%$. We analyze our
results in terms of the interaction between
the dislocation core and some isolated obstacles either single Al
placed at different positions around the glide plane or Al dimers
with different positions, orientations and bond lengths. The
dominant component of the dislocation pinning is found to be a short
range interaction of typically few Burgers vector between the
obstacles and the distinct Schokley partials of the dissociated dislocation.
This rather short range interaction is to be compared with the
partial cores which spread over few lattice
spacings.\cite{Joos1997,Szelestey2003} Our systematic study of every
dimer configuration allows us to enhance the role of the chemical
interaction between nearest solutes which yields some pinning strengths that diverge
from the simple superposition of the strain fields due to each
solute.

In addition to provide some data about the strength of
a single Ni(Al) crystal, our atomistic simulations allow us to challenge the
different versions of the statistical
theory for the SSH.\cite{Mott1948,Mott1952,Nabarro1985,Friedel1964,Haasen1979,Suzuki1991,Butt1980}
The corresponding analytical models intend to provide an
estimate of the CRSS from the elementary
interaction between a single dislocation and an isolated obstacle.
Most of the other
effects on the dislocation pinning
as the
presence of the grain boundaries, the dislocation forest and the thermal
activation of the solute motion are assumed not to play an important role.
This proves to match the conditions of our numerical simulations
which permits us to compare the theoretical predictions to the simulation data.
A common point to the different versions of the SSH theory is to have been derived in the framework of the
continuous line tension model where the dislocation is thought of as an elastic
string anchored by a single type of obstacles.
The split between the different versions of the theory stems from the various
assumptions made
on the critical string configuration and how the depinning proceeds.
For the sake of consistency, the string-obstacle interaction parameters are
determined from our atomistic study and the
statistical models end results are compared to
our direct computation of the CRSS for the Ni(Al) single crystal.
Among the theories proposed to compute
the CRSS, we show that some of them quantitatively agree
with our computations for the
concentrated solid solutions, i.e. $2$ at. $\%$  $<C_{Al}<$
$12$ at. $\%$. The concordance of the atomistic simulations and
the SSH continuous
theory confirms the plausibility of a multi-scale approach to
the plastic flow in the inhomogeneous media.

The present paper is organized as follows: In Sec. II, the atomistic
method to compute the CRSS for a screw dislocation is described
and the results are compared to the edge ones.
In Sec. III, the different analytical models for SSH are presented
and discussed in regard of our atomistic simulations. In Sec. IV, our
study is resumed and our future works announced.

\section{Molecular static computation of the solution strength}

By contrast to the phenomenological line tension approach applied to the
SSH,\cite{Foreman,Arsenault1989} the atomistic simulation allows to
capture the main physical aspect of the dislocation core by
integrating the nonlinear many-body interactions between the atoms
displaced during the dislocation course.
\begin{figure}
\noindent
\includegraphics[width= 8cm,angle=90]{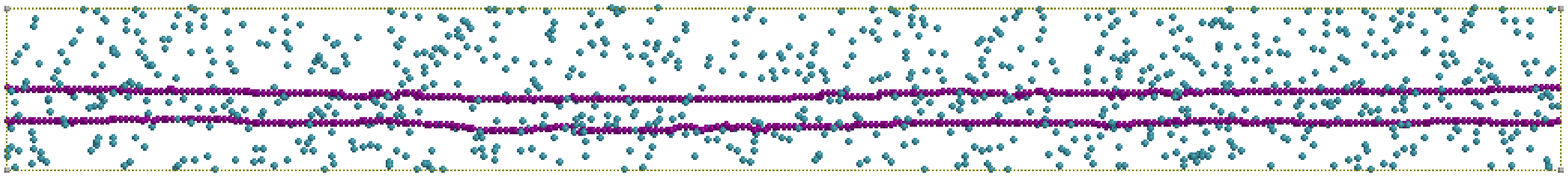}
\hspace{0.7cm}
\includegraphics[width= 8cm,angle=90]{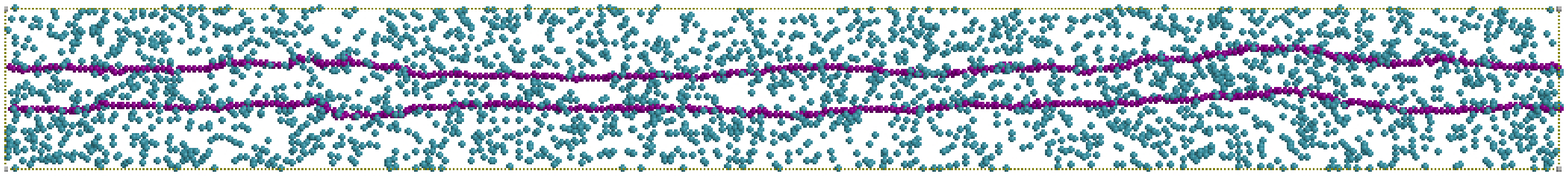}
\hspace{0.7cm}
\includegraphics[width= 8cm,angle=90]{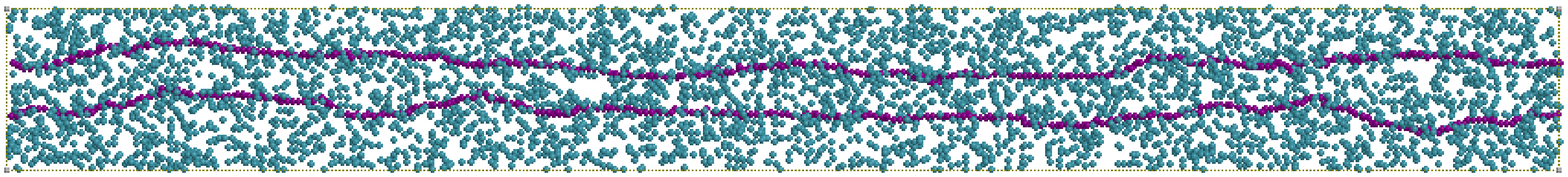}
\caption{\label{solsol} (color online) View of the $(1\bar{1}1)$
glide plane and the dissociated screw dislocation in the Ni(Al)
simulated solution at various concentrations: $C_{\text Al}=2$ at.
$\%$ (lhs), $C_{\text Al}=6$ at. $\%$ (center) and $C_{\text Al}=10$
at. $\%$ (rhs). The box size is $300$ b along Y, $32$ b along X {\color{black} and
$34$ b along Z (orthogonal to this paper sheet)}. The
Shockley partials are colored in violet and the Al atoms in blue.
The external shear stress has been fixed to one half of the CRSS.}
\begin{picture}(200,0)(0,0)
\put(-5,105){\makebox(0,0){ X}} \put(-10,130){\makebox(0,0){Y}}
\put(-18,110){\Large {\bf $\longrightarrow$}}
\put(-13,112){\begin{rotate}{90} {\Large {\bf $\longrightarrow$}}
\end{rotate}}
\end{picture}\end{figure}
In our simulations the atomic interactions are modeled through the
embedded atom method (EAM) the detail of which has been published
elsewhere.\cite{Angelo1995,Voter1987,Rodary2004} One must notice
that two typo errors must be corrected before implementing the EAM
potentials. For the Ni-Ni interaction see Ref. \onlinecite{AngeloErratum}
and for the Ni-Al interaction the coefficients $g_{\text Ni}$
and $g_{\text Al}$  must be
exchanged in Ref. \onlinecite{Rodary2004}. According to private communications
with other authors these corrections have been taken into account in
other earlier works using the same method. The simulation cell (see
Fig.\ref{solsol}) is oriented so as that the horizontal Z planes are
the $(1\bar{1}1)$ of the face centered cubic (fcc) lattice while the
Y direction corresponds to the screw Burgers vector $b=[110]a_0/2$
and the X direction is orthogonal to Z and Y and points at the
dislocation motion. The simulation box size along the directions
$i=X,Y,Z$ is denoted by $L_i$. The periodic boundary conditions are
applied along X and Y while the external applied stress $\tau_{yz}$
is produced by imposing extra forces to the atoms in the upper and
lower Z surfaces.\cite{Rodney2000} In order to form a screw
dislocation between the two $(1\bar{1}1)$ central mid-planes, the
displacement field of the elastic solution for a dissociated screw
with Burgers vector $b$ is applied to the atoms of the simulation
box. In order to compensate the Burgers vector shift at the crossing
of the boundaries along X, the corresponding periodic boundary
conditions are tilted from $b$ alongside Y. The ideal solid solution
is then formed by substituting randomly Ni atoms with Al in the
proportion fixed by the Al atomic density $C_{\text Al}$. The solute
distribution depends on the {\it seed} of the numerical
random generator. For each $C_{Al}$
ranging from 2 to 12 at. $\%$, several distributions have been generated. The Molecular
Statics (MS) simulations are performed to minimize the {\color{black} total
enthalpy} under a fixed applied shear stress. The external
applied stress is incremented by $0.3$ MPa and for each increment
the minimization procedure is repeated until it either
converges to a required precision or the dislocation starts to
glide. To minimize the size effects in the simulations of the random
solution, we chose $L_y=300b$ which allows to neglect the
interaction between an obstacle and its periodic image and to
achieve satisfactory statistics.  The same method with same inter-atomic
potentials have been employed in Ref. \onlinecite{Proville2006} for the edge dislocation.

In Fig.\ref{solsol} we reported a snapshot of three typical systems
computed for an external shear stress smaller than the CRSS
$\tau_c$. The atoms involved into the partials are recognized by
their default configurations in first neighbor positions. The Al
atoms that participate to the $(1\bar{1}1)$ planes that bound the
glide plane have also been reported on these 3 pictures. The screw
dislocation oscillates in the crystal in a way similar to the edge
one as reported earlier.\cite{Proville2006} It is worth noticing
that such a wavy profile has been observed
experimentally\cite{Saka1985} in some other type of solid solution
as Cu(Al) and Cu(Si). The dislocation does not form large bow
between well separated pinning points but rather conserves a wavy shape.
Such configurations of the dislocation impinged on a random solution
do not correspond to the one obtained within a phenomenological
elastic string model as the one used for instance in the work of Foreman and
Makin.\cite{Foreman} The reason is threefold: (i) the solution is
concentrated and thus the isolated point like obstacles are rare,
(ii) the interaction dislocation-obstacle is not a point-like force
and (iii) the pinning strength of the Al obstacle is much weaker
than what can be estimated through the elastic stress-strain field
of a Volterra dislocation.
As detailed further (see Tab.\ref{Table1}, the pinning
strength of an isolated Al on the screw dislocation is about one
hundredth of the line tension meanwhile the elastic theory in a very
first order version would predict one order larger.

\begin{figure}
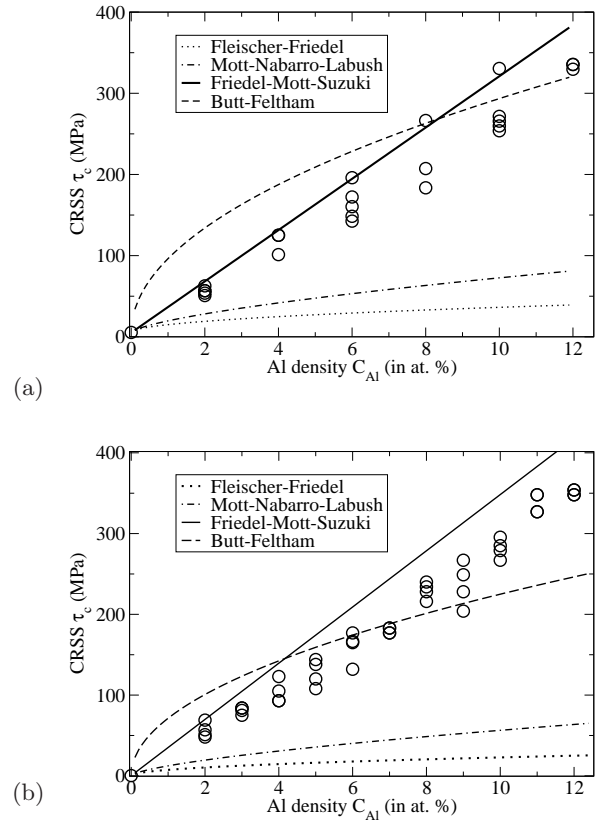

\noindent
\includegraphics[width= 7cm]{NiAl_SSH_Vis.eps}\\
\vspace{0.8cm}
\includegraphics[width= 7cm]{NiAl_SSH_Coin.eps}
\caption{\label{ssh_screw} Variation of the critical resolved shear
stress $\tau_c$ for a screw dislocation (a) and for an edge
dislocation (b) against the Al concentration $C_{\text Al}$ computed
from the MS simulations with different Al random distributions
(symbols). The estimations made with the analytical models have also been
reported: Fleischer-Friedel\cite{FleischerHibbard1963} (dotted line), Friedel-Mott-Suzuki\cite{Friedel1964} (full line), Mott-Nabarro-Labusch\cite{Nabarro1985}
(dot-dashed line) and Butt-Feltham\cite{Butt1980} (dashed line).}
\begin{picture}(200,0)(0,0)
\put(-15,270){\makebox(0,0){(a)}} \put(-15,117){\makebox(0,0){(b)}}
\end{picture}
\end{figure}

In our MS simulations when the dislocation moves freely, we let it
glide for ten passages in the simulation box in order to probe
possible stronger pinning configurations created by the relative
displacement of one Burgers vector $b$ at each passage between the
plane above and below the glide plane. Since $L_x \approx 32$ b, we
assume that when the dislocation has glided over $10\times L_x$ we
reach the maximum of the applied shear stress for the considered random distribution.
Such a glide distance has the same order as
the mean free path of a gliding dislocation through the forest dislocations in a
metallic polycrystal where the dislocation density may
reach $10^{9}$ cm$^{-2}$. In most of our computations the
dislocation does not encounter a new pinning configuration after 4
passages in the simulation box. For each concentration, the
calculations of the CRSS, $\tau_c$ from various distributions have
been reported in Fig.\ref{ssh_screw} (a) where each open symbol
corresponds to a different random distribution. The dispersion on
the measure of $\tau_c$ is related to the finite size effect along
Y. The choice of the suitable $L_y$ results from a
compromise between the computational load and the statistics.

The
order of magnitude of the solution strengthening is about 30 MPa per
atomic percent of Al which is noteworthily similar to the edge one,
obtained from Ref. \onlinecite{Proville2006} and reported in
Fig.\ref{ssh_screw} (b) for further comparison with the analytical
models.
Such a result contrasts seriously with the classical calculations
based on the elastic theory of dislocation.\cite{Arsenault1989}
However this can be fairly understood from the analysis of the
interaction between the dislocation and a single obstacle at the
atomistic level. We carried out the same type of MS
simulations as for the random solid solution except that only one
isolated Al atom is placed in a simulation cell of pure Ni. Then the
external shear stress $\tau_{yz}$ is incremented from zero to
$\tau_m$ at which the dislocation liberates. Since the simulation
cell is periodic along Y, the obstacle and its periodic images form a
regular array of obstacles separated by $L_y$. The balance between
the Peach-Kohler force and the obstacle pinning leads to
$f_m=(\tau_m-\tau_p) b L_y$ where $\tau_p$ is the screw Peierls stress and $b$ is the Burger vector of the whole
dislocation. Because of the nonlinearity of the atomic
interactions, we found that the pinning strength of an isolated
obstacle depends on its place above or below the glide plane as well
as on which partial is concerned. Our results are
reported on the first lines of Tab.\ref{Table1} where the
corresponding strength $f_m$ has been normalized by the constant
$\mu b^2$ to provide the standard pinning coefficient denoted by
$\alpha$. We choose $\mu=74600$ MPa, the shear modulus
for the Ni $(111)$ planes $[c_{11}-c_{12}+c_{44}]/3$.
Although the normalization has been realized
with the value of $\mu b^2$ for pure Ni, one must bear in mind
that this normalization is a conventional way to present the
result.
In Tab.\ref{Table1} the single obstacle denoted by
(a1) corresponds to an isolated Al placed in the $(1\bar{1}1)$ plane
situated just above the glide plane while (b1) is for an Al which
participates in the $(1\bar{1}1)$ plane just below the glide plane. The
pinning strength of a single Al atom is found to have the same
magnitude as for the edge dislocation\cite{Proville2006} which
confirms our results about the same solid solution CRSS for both edge and
screw dislocations. The
strengths of  some isolated Al have also been reported in Tab.\ref{Table1},
for the farther $(1\bar{1}1)$ planes. For the second nearest one,
the pinning forces are referenced by (a2) and
(b2) for the Al above and below the glide plane, respectively.
For the third nearest planes,
the pinning forces are arranged in same order and referenced by (a3) and (b3).

\begin{figure}
\noindent
\includegraphics[width=7cm]{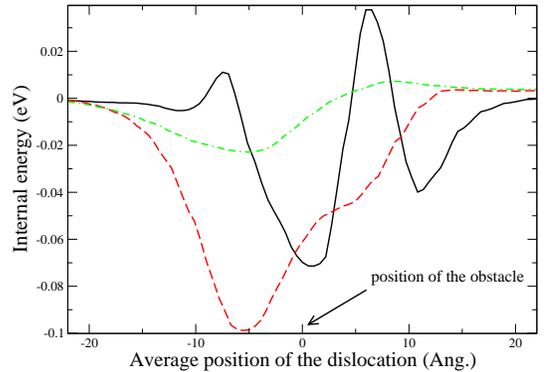}
\caption{\label{Poten} The internal potential energy {\color{black} for a
dissociated screw dislocation} against the
dislocation obstacle distance. The obstacle is an isolated Al in the
$(1\bar{1}1)$ planes above the glide plane: the first plane (full line), the second plane
 (dashed line) and the third one (dot-dashed).}
\end{figure}

{\color{black}
On the condition that the dislocation core distance to the obstacle remains larger than
the core extent, it is possible to analyze the interaction in terms of the continuous
elastic theory. We intend applying such an approach for the obstacles situated in the third $(1\bar{1}1)$ plane
from the glide plane.}
For that case, the potential landscape of the {\color{black} dissociated screw} dislocation (dot-dashed line in Fig. \ref{Poten})
has the form $\beta z[1/(z^2+(x+d)^2)-1/(z^2+x^2)]$, i.e. the one for two opposite edge dislocations
separated from $d$, the dissociation distance.\cite{Friedel1964,Suzuki1991,Hirth1982}
According to the elastic theory, the interaction
pre-factor $\beta$ is given by $\frac{\mu b^*}{3\pi}\frac{1+\nu}{1-\nu}(\Delta V)$ where $\nu$ is the Ni Poisson
coefficient, $\Delta V$ is the atomic volume variation due to the Al impurity in Ni.
If one neglects the interaction between the obstacle and the farthest partial,
the corresponding pinning coefficient for a single partial is thus
$\alpha=\frac{1+\nu}{1-\nu}(\Delta V)b^* \sqrt{3}/(8\pi b^2 Z^2)$.
The distance Z between the third $(1\bar{1}1)$ plane and the glide plane is $Z=5b/\sqrt{6}$. This corresponds to
two and a half of the inter-plane distance along the $[111]$ direction.
To provide an estimate of the pinning force we choose $b^*=b/\sqrt{12}$
which is the edge component of a perfect Shockley partial.
The volume variation can be estimated according to a method described in
the textbook,\cite{Friedel1964b}
$\Delta V= 3v_{Ni} a_0^{-1} da_0/dC_{Al}$, where the Vegard's law for Ni(Al) is used to express
$a_0^{-1} da_0/dC_{Al} = 0.0763$. Then one found $\alpha=0.0015$ which is of same order though still larger than
what has been found in our simulations (see
lines (a3-b3) in Tab.\ref{Table1}). If instead of $b^*=b/\sqrt{12}$, one chooses the effective Burgers vector
computed through a Peierls-Nabarro method\cite{Szelestey2003} then $\alpha=0.00083$ which agrees better.
{\color{black} A more quantitative
study would require the account for the modulus misfit.}
This comparison allows to
emphasize that as
a consequence of the edge Burgers components of both partials
the size effect dominates the modulus misfit effect often
invoked in the analysis of the screw-impurity interaction.
Since the partials have some
Burgers vectors that are not purely screw, a hydrostatic stress
field is localized around the partial cores\cite{Wen2005} which
renders the partials of a screw similar to the ones of an edge, at
the atomic scale. Far away from the center of mass of the
dislocation, since the edge components of the leading and the
trailing screw partials are opposite, the hydrostatic stress field
of both partials annihilate each other and the elastic theory
prediction for a purely screw dislocation is recovered.
In the first and second $(1\bar{1}1)$ planes, the shape of the interaction potential (full and dashed lines
in Fig. \ref{Poten}) is imposed by the nonlinear
atomic interaction involved during the passage of the impurity
through the core of the dislocation.
{\color{black} Concerning the pinning strength on a single Al atom,
we distinguish a common trend for screw and edge dislocations:\cite{Proville2006}
The anharmonicity
enhances the pinning strength in the compressive regions in regard of the tensile ones. In our simulation cell, the compressive region of the edge dislocation is situated
above the glide plane whereas for the screw 2 compressive regions can be distinguished, e.g.  below the glide plane for the leading partial and above for the trailing partial.
The obstacle labeled a1 (b1) in Tab. \ref{Table1} visits the compressive region
when it crosses the trailing (leading) partial. In agreement, the stronger pinning strength is found for the trailing (leading) partial.
Furthermore
the comparison of the
pinning strengths in the separated compressive regions
shows that the trailing partial is anchored more strongly than the leading one. The same trend
can be noted for the tensile regions.
Worthily the previous remarks are consistent with the results on the edge dislocation.\cite{Proville2006}}
The latter trend finds some substantiations into the fact that
the more stable position of the Al solute is inside the
stacking fault ribbon, as can be seen in Fig. \ref{Poten}.
The potential energy measured with respect to the isolated Al far away from the
dislocation core should yield a diffusion current toward
the dislocation stacking fault. This is a Suzuki-type effect which
operates on both types of dislocation.
This could lead to a classical
dynamic strain aging of the fcc substitutional
alloys.\cite{Chen1992} In our computation the diffusion is frozen
and such an effect is thus disregarded as it is indeed in the
SSH statistical models discussed below. Another consequence of the
absence of the solute diffusion is that there is no short range ordering even for the Al concentrated solid solutions.
It must be noticed that
the interaction potential with the nearest obstacles (see Fig.\ref{Poten})  is
far from the simple shape assumed in some more phenomenological
approaches\cite{Nabarro1985} and even from an elastic
theory\cite{Arsenault1989,Gremaud1999} which neglects the
dislocation core geometry.
As it is required for the construction of a statistical
model, some of the details of the atomic scale must be left off
and one must retain only the essential quantities that are
identified as important. According to the SSH theory, an obstacle
can be characterized by an interaction range $w$ and a pinning force $f_m$,
or equivalently a binding energy $U$.
We thus attempt to estimate such quantities and to that purpose we chose
arbitrarily a rounding method.
We distinguish only two different
cases depending on the obstacle position with respect to the leading
and the trailing partial. For the pinning strength $\alpha$ we choose
the maximum of the interaction force.
The definition of the
interaction range $w$ requires a more tactful treatment since in principle the short
range obstacle-core interaction is superposed to a long range
elastic one and further the interaction potential is not symmetric
for an obstacle position ahead and behind the partials, mainly
because of the presence of a stacking fault. Around the absolute
force maximum, we measure w as the shortest distance to which the force vanishes or falls
to a local minimum. We choose to neglect the
variations of the force over that limit. We concede that such a
point can be discussed and must be kept in mind for further
discussion on the analytical models. We emphasize that we have been
primarily concerned with finding a manner to estimate one of the key
parameters of the theory, $w$ and that there is no well prescribed
theoretical way for that. We believe that the SSH theory should be developped
to address specifically the case of
a dissociated dislocation with an
interaction potential similar to the one reported in Fig.\ref{Poten} rather than the
standard Gaussian-like potentials.

In the present study, when
the interaction potential shows a force maximum near the position of one of the partials
we ascribe the force to this partial. This is a conventional way of arranging the numerous
obstacle forces since the two partials are actually interacting through the stacking fault.
Further when the obstacle is attractive for the leading partial and repulsive for the trailing one, the
force fields overlap and it may then be difficult to identify with accuracy the interaction range.
When it has not been possible to separate different force maximum, we reported
the corresponding pinning strength
as being exerted on the trailing one (see (b2-b3) lines in Tab.\ref{Table1}).
\begin{table}
\caption{\label{Table1} Summary of different pinning obstacles for
both the leading (subscript $l$) and the trailing partials
(subscript $t$) of the screw dislocation, their pinning force
$\alpha$ (normalized by $\mu b^2$) and their force range $w$.}
\begin{ruledtabular}
\begin{tabular}{cccc}
 Nature & text Ref.  &  $\alpha_l$ and $w_l$ &
 $\alpha_t$ and $w_t$ \\
\hline single & (a1) & $0.0020$/ $1.14b$ &  $0.0111$/ $0.88b$ \\
1st planes& (b1) & $0.0088$/ $1.05b$  & $0.0059$/ $2.20b$ \\
\hline
single & (a2) & $0.0036$/ $1.41b$ &  $0.0032$/ $1.69b$ \\
2nd planes& (b2) & -  & $0.0053$/ $1.52b$ \\
\hline
single & (a3) & - &  $0.0011$/ $2.8b$ \\
3rd planes& (b3) & -  & $0.0009$/ $2.85b$ \\
\hline
1st neighbor & (c) & $0.0022$/ $0.77b$  & $0.0264$/ $2.19b$  \\
non-crossing pair & (d) & $0.0006$/ $0.1b$  &  $0.0139$/ $1.49b$ \\
& (e)   & $0.0033$/ $1.03b$ &  $0.0168$/ $1.14b$ \\
& (f)   & $0.0079$/ $0.72b$  &  $0.0086$/ $2.82b$ \\
& (g)    & $0.0207$/ $1.8b$  & $0.0068$/ $2.03b$ \\
& (h)    & $0.0153$/ $1.05b$          & $0.0106$/ $2.33b$ \\
\hline
3th neighbor     & (i)    & $0.0094$/ $1.24b$          & $0.0147$/ $2.45b$ \\
crossing  pair   & (j)    & $0.0026$/ $0.43b$          & $0.0206$/
$2.09b$  \\
\hline
2nd neighbor     & (k)    & $0.0137$/ $1.24b$          & $0.0142$/ $2.45b$ \\
         crossing  pair    & (l)    & $0.0184$/ $2.87b$          & $0.0224$/ $2.38b$ \\
                           & (m)    & $0.0096$/ $1.09b$          & $0.0146$/ $1.07b$ \\ \hline
                1st neighbor     & (n)    &   -        & $0.0157$/ $1.79b$ \\
          crossing  pair   & (o)    & $0.0171$/ $1.87b$          &  $0.0035$/$1.14b$       \\
                           & (p)    & $0.0218$/ $2.77b$          & $0.0161$/ $1.37b$ \\ \hline
\end{tabular}
\end{ruledtabular}
\end{table}
In the fcc symmetry, for the Al concentration we are concerned with,
the density of the Al dimers has same order as the density of single
Al, i.e. $C_{Al}\approx nC_{Al}^2$ where $n=12$ is the number of
nearest fcc neighbors. Above $C_{Al}=1/n$ the number of isolated Al
vanishes in average. It is thus of some interest to study the
pinning strength of the Al dimers that might be expected to play a role
on the SSH because of the alloy ordering energy. Among the different configurations of pairs, we selected
those with a distance between solute atoms corresponding to first,
second and some of the third neighbors in the fcc lattice. Either the screw
dislocation interacts with preexisting Al dimers
referred to as (c-h) in Tab.\ref{Table1} or the dislocation passage
modifies the Al-Al bond crossing the glide plane (i-p)
in Tab.\ref{Table1}.
For the non-crossing pairs, the (c-e) configurations correspond to the planar dimer situated above the glide plane
whereas the (f-h) are below.
The directions of the dimer bond before the
dislocation passage are $[011]$ (c) and (f), $[10\bar{1}]$ (d) and (g),
$[110]$ (e) and (h), $[011]$. The dimers that cross the glide plane are
oriented in the direction $[721]$ (i), $[211]$ (j), $[212]$ (k), $[\bar{1}\bar{2}2]$ (l),
$[2\bar{2}\bar{1}]$ (m), $[4\bar{1}{1}]$ (n), $[1\bar{1}{4}]$ (o) and $[1\bar{4}{1}]$ (p).
Our results on the pinning
strength and the interaction range are reported on the corresponding
lines in Tab.\ref{Table1}.
Besides the variable geometry of the crossing pairs,
it is worth noting that the pinning
strength of a non-crossing dimer does not correspond to the simple superposition of
the strength of 2 isolated Al that would be placed at the reticular
sites occupied by the dimer. The resulting strength can be even smaller than an isolated
Al as seen from the comparison
between the single (b1) and the first neighbor dimer (f).
It must be also remarked that some of the dimer strengths can be larger than two isolated Al
at the same position, e.g. (b1) and  (g) in Tab.\ref{Table1}.
The second neighbor Al dimers could be expected to be more stable than others with regard to the
$L1_2$ ordering trend of the Ni(Al) alloy.
Such an expectation is not reflected by the dimer pinning strengths which are not significantly larger to
break the second neighbor pairs although
the interatomic potential between Ni and Al particles
was adjusted to fit the corresponding order energy.\cite{Rodary2004}
The forces reported on lines (k-m) are not particularly stronger than
those concerning the first and third neighbor dimers.
Finally if one considers the whole set of
the dimer configurations, the pinning strengths
for a screw dislocation are not smaller than for the edge ones reported in Ref. \onlinecite{Proville2006}.

\section{Analytical models for computing the solution CRSS}
The MS simulations allowed us to perform a direct computation of the
CRSS at the atomic scale. In the past, different statistical
theories have been proposed to evaluate the CRSS of solid solutions. The common assumption of
the various analytical models is that the CRSS can be derived only from a unique
interaction of the dislocation with an isolated impurity in the glide plane.
Although these statistical models were primary devoted to the computation of the CRSS for an edge dislocation,
there is no theoretical argument against the
application of these models to a screw dislocation provided that one
is able to quantify the model input parameters.
In order to compare consistently
the model predictions to our data on the SSH, the model parameters
are determined from the average of the single Al pinning configurations
(lines (a1) and (b1) in Tab.\ref{Table1}).
We obtained a
single valued obstacle strength $f_m=\bar{\alpha} \mu b^2$ where
$\bar{\alpha}=0.007$ is the mean pinning coefficient for
the isolated Al situated in the
nearest $(1\bar{1}1)$ planes.
We also computed the typical interaction range as an
average over the different isolated obstacle configurations ${\bar w}=1.3 b$.
Further, since in principle the models apply to an undissociated
dislocation, the leading and the trailing partials are assumed to be
tightly bound.
In the different models, the end formula is usually presented for
a theoretical square lattice which leads to an atomic area of $s=b^2$.
In our system, this is to be changed for $s=b^2\sqrt{3}/2$,\footnote{This corrects a mistake
on $s$ introduced in
Ref. \cite{Proville2006} with no consequence on the end results in same reference} i.e. the
atomic area in the glide plane.

Another quantity required before going into the details of the
models is the dislocation stiffness. In our simulation cell, we
computed the dislocation line energy as the difference between the
whole atomic potential energy of a simulation cell with a
dislocation and the one of a perfect single crystal with same
geometry and same atom number.
{\color{black} It was then possible to
obtain the energy per unit length $E_L$ of the straight dislocation in pure
Ni. One must emphasize that this computation implies the stacking
fault energy which is not modeled in the classical estimation of the
dislocation energy, i.e., $E_L= \mu b^{2} /2$.
For the edge
dislocation, we obtained $E_{Le}=0.95 E_L$
while for the screw $E_{Ls}=0.61E_L$. These two quantities depend on the simulation cell
geometry ($L_x$,$L_z$) in agreement with the logarithmic law derived from
elastic theory. }
Since in the
Ni(Al) solid solution (see Fig.\ref{solsol}) the dislocation length
depends on the Al distribution, it is not possible to compute the
dislocation line energy as done in a pure material where the
dislocation is straight. However, one can expect reasonably that $E_L$ should
vary as the elastic shear modulus $\mu$ and the lattice parameter
$b$, according to the Vegard's law which leads to a variation of roughly  1 at. $\%$ of $C_{Al}$.
Since such a variation of the dislocation stiffness has not been considered in the SSH theory
we shall not account further for such a
dependency although for the concentrated solutions it would be worth accounting for it.
To obtain the dislocation line tension we apply a
well known result from the isotropic elastic theory\cite{Hirth1982b} which
tells us that for the screw $\Gamma_s=E_{Ls}(1+\nu)/(1-\nu)$ while for the
edge $\Gamma_e=E_{Le}(1-2\nu)$.
{\color{black} We note that the line tension increase with the cell
dimensions will impact the dislocation roughness through the obstacles distribution
as it is predicted by some of the following models.}

\subsection{Fleischer-Friedel theory}

One of the classical models to evaluate the strengthening of the
solid solution is referred as to the Fleischer-Friedel (FF)
model.\cite{FleischerHibbard1963} The model's main assumption is that
the dislocation line forms some bows\cite{Foreman} between isolated point-like
obstacles randomly distributed. The bow shape assumption contrasts
with what can be seen from our simulations in Fig.\ref{solsol}.
The critical angle associated with the impurity
strength is $\bar{\alpha}_s=f_m/2\Gamma_s$.
The Friedel
length is $L_F=\sqrt{2s\Gamma_s/cf_m}$ where $c$ is the obstacle
concentration which corresponds to
$c=4C_{Al}$ to account for the different configurations, i.e. above and
below the glide plane and for both partials (Tab.\ref{Table1}).
The CRSS is then given by the balance
between the Peach-Kohler force and a meanfield regular array of
obstacle $\tau_c b L_F = f_m$. The FF end formula reads:
\begin{equation}
\label{5b1}
  \tau_c=\frac{f_{m}^{3/2}\sqrt{c}}{b\sqrt{2s\Gamma_{s}}}.
\end{equation}
The corresponding plot of $\tau_c$ against the Al density is reported in Fig.\ref{ssh_screw} (a) and (b)
for the screw and the edge dislocation, respectively. The FF theory is found to underestimate the CRSS of our
simulations. Foreman and
Makin\cite{Foreman,Rodney2006} showed that the Eq.\ref{5b1} overestimates by
about $10$ $\%$ the true CRSS for some perfect point-like obstacle
random distribution so that the discrepancy between our simulation
and the FF model predictions could not be compensated by such a
correction.  It is compelling that the FF model is not suitable
to compute the CRSS in the concentrated random Ni(Al) solution.
One must be reminded however that the FF model proved reliable\cite{Foreman}
for some stronger obstacles as precipitates and for smaller densities. It would be interesting to
study by MS the lower densities but then the finite size sampling of
the Al distribution limits the statistics.

\subsection{Mott-Nabarro-Labusch theory}
The Mott-Nabarro-Labusch (MNL) theory has been built from different
contributions.\cite{Mott1948,Mott1952,Nabarro1977,Nabarro1985,Labusch1970,Labusch1972}
The solute dislocation interaction is assumed to be of a finite
range $\bar{w}$ without presuming of the attractive or the repulsive
character of the interaction. We introduce the parameter $v=2\bar{w}$ which
agrees with the definition of Nabarro for the interaction range. The
dislocation is still idealized by a perfect elastic string and its
configuration is assumed to be a quasi-straight line which
 interacts with several solute atoms. Alongside the dislocation core, in a
ribbon of length $2L$ and width $2v$, the number of atomic sites in
the planes contiguous to the glide plane is $8vL/s$. The ribbon
extent allows to account for the attractive and repulsive parts of
the interaction. Inside this ribbon a counting of the obstacle gives in average
$2n=8vLc/s$. The segment is sustained to a
mean restoring force which depends on the stiffness of the string.
Mott\cite{Mott1952} showed that this force could be written as
$2Lf_m^2 x/L'v^2\Gamma_s$ where $x$ is the segment mean position and
$L'$ is the mean distance between two obstacles situated along the
segment $2L$. The distance $L'$ is fixed by $4L'vc/s=1$. Nabarro
assumed\cite{Nabarro1985} that the characteristic length $L$ could be identified as
the distance above which
 the string Green's function vanishes:
 $L=(L' v \Gamma_s/\sqrt{2}f_m)^{2/3}$. Following Mott and
Nabarro,\cite{Mott1948} the counting of the obstacles situated in
front and behind the segment $2L$ leads to a total number of
interactions $2n$ which the  average force is $\pm f_m/2$. These
interactions yield a maximum fluctuation of $\sqrt{2n} f_m/2$
which must be equated to the external Peach-Kohler force
$2Lb\tau_c$ to obtain
the CRSS:
\begin{equation}
\label{5b3} \tau_{c}=\left(\frac{c^{2}v
f_{m}^{4}}{b^{3}s^{2}\Gamma_{s}}\right)^{1/3}.
\end{equation}
Here the obstacles above and below the glide plane have been
accounted for. If one assumes that the two partials are bound, one
must replace c with $2C_{Al}$. The plot of the corresponding CRSS has
been reported in Fig.\ref{ssh_screw} and it is found to
underestimate the strengthening for the concentrated solutions.
The same results hold for the edge dislocation. It is noteworthy that
the MNL model is commonly thought to provide a good description for the high densities.
In the case of our Ni(Al) solution, our comparison contrasts with such a belief.
However, as for the FF model at smaller densities it is not excluded that the agreement could be recovered.
An extended version of the MNL model has been proposed
elsewhere\cite{Proville2006} to tentatively account for the
different types of obstacles, i.e. isolated and clusters with different
pinning strength $f_m$ and interaction range $w$. Although this
model gave us some satisfactory results for both the edge and the screw dislocation further developments are
required to be fully consistent on the theoretical treatment.

\subsection{Butt-Feltham theory}

In the Butt-Feltham (BF) theory,\cite{Butt1980,Butt1981,Butt1993} an undissociated dislocation
liberates from a pinning configuration by nucleation of a bulge.
This critical bulge can be approximated by a triangle shape of
height $W$ which corresponds to the saddle point energy required to
unzip the whole dislocation. According to Feltham $W$
is estimated from the dislocation core radius extent and thus for a close-packed metals
$W\approx 3b$. As other previous models in the BF theory it is proposed to
relate the CRSS to the in-plane obstacle density. Along the
dislocation line the inter-obstacle distance is roughly
$b/\sqrt{c}$. For a quantitative comparison we distinguish the
mean distance in function of the different geometry for
the edge
and the screw dislocation: $\lambda=\sqrt{2}b/\sqrt{c}$
and $\lambda=b/\sqrt{c}$, respectively. The enthalpy required to
form a bulge is : $ H=U L/\lambda + 2W^2\Gamma/L -\tau bW L/2$ where
$U$ is the binding energy and $L$ is the bulge extent along the
dislocation line. In the expression for $H$, one recognizes the binding energy,
the elastic cost for the bulge and the work of the Peach-Kohler force.
The
bulge curvature is determined by the external stress and thus for a certain W we deduce
the corresponding bulge extent
$L=\sqrt{8W\Gamma/\tau b}$. At the critical configuration, the enthalpy cancels
which gives
the end result:
\begin{equation}\label{BF}
\tau_c=\frac{4U}{bW \lambda}.
\end{equation}
We assume that the binding energy can be estimated roughly as $U=f_m \bar{w}$ and we account for
the different obstacle configurations by $c=4C_{Al}$.
The corresponding plot has been reported in Figs.\ref{ssh_screw} (a) and \ref{ssh_screw} (b)
for both edge and screw.
Although the predicted CRSS is quantitatively of same order as the MS simulation data, the BF theory predicts
a $\tau_c$ in $\sqrt{C_{Al}}$ which underestimates the hardening rate of our simulations. It should be mentioned
that Butt and Feltham also derived different power laws according to the variation of the bulge height $W$.
With such an extension the BF theory prediction may agree better with the MS data.

\subsection{Friedel-Mott-Suzuki theory}

Another theory proposed by Friedel, Mott and Suzuki (FMS) in their
text books\cite{Friedel1964,Suzuki1991} considers the few atoms
around an ideal undissociated dislocation line. This theory allows
the dislocation to take locally much larger curvatures than in the
FF model. The first effect of this pinning of the dislocation is to
give the dislocation a ``zigzag'' shape. The interaction is
characterized by a binding energy $U$ which we approach by $f_m \bar{w}$
valid for a linearized force. We
briefly recall the model derivation. The amplitude of the zigzag in
the glide plane is denoted by $W$ and its wavelength
alongside the dislocation line is denoted by $L$. According to
Friedel\cite{Friedel1964} the number of solute contained in the
glide plane area $WL/2$ is given by the inverse of the surface
density, i.e. $WL/2= s/c$. However a careful counting leads us to
$WL=s/c$ since the construction of a regular two-dimensional lattice
where the obstacles are separated by $L$ in the Y direction and $W$
in the $X$ direction leads us to $WL=s/c$ which agrees with Suzuki's derivation.\cite{Suzuki1991} The binding energy per unit
of length is $E_1=2f_m\bar{w}/L$. The line tension energy involved in the
zigzag is given by $E_2=\Gamma(\sqrt{(W^2+L^2/4)}-L/2)/(L/2) \approx
2\Gamma W^2/L^2$ at first order in $W/L$. By minimizing the
difference $E_2-E_1$ one obtains the optimal value of the zigzag
$W=(f_m \bar{w} s/4\Gamma c)^{1/3}$ that fixes the line shape and thus
$E_1= 2\bar{w}Wc/s$ and $E_2= 4\Gamma c^2 W^4/s^2$. The application of
external shear stress deforms the dislocation. It has been assumed
that the effect of the external shear stress is to unzip the line
from the obstacles at the bottom of the zigzag, where the line
tension exerts a maximum force in the glide direction. Considering
that the maximum of the potential energy is a straight line ($W=0$),
bound to the obstacle situated at the top of the former zigzag, the energy difference
between this maximum and the zigzag configuration is $(-E_1/2 )-(E_2 -
E_1)$. Multiplied by the wavelength $L$ this gives the work that
must be provided by the external shear stress to overcome the
pinning barrier. The area that is comprised between the two
configurations of the line, zigzag and straight is $W L/2$, so the
stress work is $\tau_c b W L/2$ which must equal $( E_1/2-E_2)L$. The end result for
$\tau_c$ is:
\begin{eqnarray}
\tau_c&=&\frac{f_m \bar{w} c}{sb}.
\end{eqnarray}
Noteworthily this expression for the CRSS does not depend on the
line tension in contrast to the other models. To apply the
theory to our system, the input parameters are estimated similarly
and we assumed that $c=4C_{\text Al}$.
In Fig.\ref{ssh_screw}(a) and
(b), it is remarkable that the $\tau_c$ linear dependency in $C_{Al}$ is much closer from
the simulations than a fractional power law.
The FMS assumption of a zigzag shape
is  closer from the wavy profile of the dislocation seen
in Fig.\ref{solsol}. The agreement between our data and the
theory is better for the screw (Fig.\ref{ssh_screw}(a))
than for the edge (Fig.\ref{ssh_screw}(b)) since for
the latter the FMS model slightly overestimates the CRSS.
With the
force model parameters for a screw dislocation, the amplitude of
the zigzag is $W=0.43 b$ at $C_{Al}=2$ at. $\%$ and $W=0.25 b$ at
$C_{Al}=10$ at. $\%$. Besides the fact that such an amplitude is
smaller than what can be depicted in Fig.\ref{solsol}, the model
predicts $W$ to decrease with $C_{Al}$. We studied the
dislocation roughness and found the opposite trend.
In view of the discrepancy on the edge CRSS, our procedure to compute the model parameters
from our atomistic data can be discussed.
To improve this transfer from the atomic scale to a continuous theory, we emphasize that some
improvements of the model would be required to account for the dissociation of the dislocation
core and the various types of obstacles.




\section{Summary and perspectives}

The aim of the present paper was to extend the study of CRSS for
a solid solution Ni(Al) to a screw dislocation. We carried out two
types of computation: (i) the CRSS of a random distribution of
solutes at different concentration $C_{Al}$ and (ii) the pinning
strength of some different obstacles, isolated Al or dimers. We
found that the CRSS and the pinning strengths of the screw
dislocation are of the same order as the edge ones
studied elsewhere.\cite{Proville2006} Such a result could not be
expected from a first order elastic theory which conventionally
ascribes the screw pinning to a modulus misfit effect in contrast to
the dominant size effect in the edge case. For the
isolated Al impurities, this result has been clearly identified as a
consequence of the dissociation of the dislocation core and the
edge components of the Shockley partials cores. The Al
clusters pinning forces have been shown not to derive from the linear
superposition of the isolated obstacle strain field.

Another issue of the present paper was to tentatively apply some
analytical models to compute the CRSS. To that purpose, the elementary
input parameters of the models were compiled from our atomistic data. The analytical models
that we considered only account for the isolated foreign atoms situated in the planes
that bound the glide plane.
The Friedel-Mott-Suzuki theory provides us a better agreement for the Ni(Al) model solid solution.
The estimation of the interaction range between the dislocation and the obstacle
is however to be questioned because
of the presence of the stacking fault and other nearest obstacles.
Our work emphasizes that the SSH
theory has to be developed to integrate more of the atomistic
details of the dislocation solute interaction.
We believe that
the accuracy of the SSH theory in the fcc solid solution
requires to integrate the dissociation of the dislocation as well as the possibility
for cluster formation in the concentrated solution.
The farther obstacles
as those situated in the second and third $(1\bar{1}1)$ planes were found to yield
a pinning strength that is not negligible in comparison to the solute atoms that bound the glide plane.
We also tested other SSH theories involving
mixing laws as the Pythagorean one \cite{Hiratani} but they are
essentially devoted to model the random distribution of
point-like obstacles and finally did not allow us to obtain a better
agreement.

To predict
quantitatively the CRSS in a variety of materials, we believe that an extended use of the
empirical interatomic potentials is not feasible at the present state of our skill in
developing such potentials. The recent progress of the first principles calculations
for alloys and dislocations\cite{Gornostyrev2007,Lisa2008} allows us to expect that the pinning strength
could be computed with a better accuracy. However it is difficult to imagine that the statistics could be
also studied with such methods because of the associated computational load. The
effective interatomic potential can thus be employed to explore the frontier between the atomic details and
the SSH statistics as we attempted in the present paper.
In the near future, we shall propose to extend our work to the Al(Mg) solid solution.
Our motivation rests on
the fact that the interatomic potentials were proved
physically reliable for the study of the alloy plasticity\cite{Olmsted2005,Curtin2006}
and the physical properties of the Al(Mg) solid solution differ from the Ni(Al) ones, e.g. a larger
size effect and a higher stacking fault energy.
The temperature effect on the SSH and
therefore the dynamics of a dislocation in a disordered media is an
eagerly difficult problem at the atomic scale because of the
time limit of the molecular dynamics simulations.
Moreover some puzzling problems of physics can be anticipated: (i) at the low
temperature the inertial effect leads to a loss of
strength,\cite{Isaac1988} (ii) at higher temperature the activation
of the diffusion yields a dynamic aging\cite{Curtin2006} and (iii)
for the intermediate temperature an odd athermal plateau on the
yield stress has been clearly identified experimentally for various
solutions and remains difficult to interpret through a simple
theory.\cite{Nabarro1985,Suzukiii}
We keep scrutinizing the works about those topics which suffer from a lack of modern investigations.


\end{document}